\documentstyle[12pt]{article} \textwidth=15cm \textheight=22cm

\begin{document}
\begin{center}
{\large \bf{Perturbative Approach for Non Renormalizable Theories.}}
\\
\medskip
{\bf J. Gegelia, G. Japaridze, N. Kiknadze and K. Turashvili} \\
\smallskip
{\it High Energy Physics Institute Of TSU, \\
University street 9,
Tbilisi 380086, Republic Of Georgia.} \\
\medskip
\end{center}
\begin{abstract}
Renormalization procedure is generalized to be applicable for
non renormalizable theories.
It is shown that introduction of an extra expansion parameter allows
to get rid of divergences and to express physical quantities
as series of finite number of interdependent expansion parameters.
Suggested method is applied to quantum (Einstein's) gravity.
\end{abstract}
\medskip
\section{Introduction}

Existence of  divergences is one of the basic problems
of quantum field theories (QFT). The renormalization procedure handles
these divergences only for some class --- renormalizable theories.
Although it is not {\it a priory} clear that non renormalizable theories
lack physical significance. Moreover, in spite of the fact that most of
the fundamental interactions are described by renormalizable QFT-s, the
problem of the quantum gravity is still open --- while Einstein's
classical theory of gravity has substantial success, the
corresponding quantum theory is non renormalizable.

We share the opinion that the renormalizability is just a technical
requirement and it has nothing to do with the physical content of
the QFT \cite{1}. A lot of people believe that in meaningful theories
divergences arise due to the perturbative expansion.
It was noted in various papers and various contexts \cite{1},
\cite{2}.
Of course not all of the non renormalizable theories are meaningful.
But the same is true for the renormalizable ones.
E.g. the scalar $\phi^3$ theory is renormalizable for space-time
dimensions up to six \cite{3}, but has spectrum unbound from below. On
the other hand there exist non renormalizable theories which can be
handled in some other approach (e.g. the four-fermion interaction in
$(2+1)$ dimensions is non renormalizable if the conventional
renormalization procedure is applied but can be renormalized after
performing $1/N$ expansion with $N$ being the number of
flavours \cite{4}).

Below we are going to present a method of extracting physical
information out of the perturbative series of non renormalizable
theories. For renormalizable ones it just coincides with
the usual renormalization procedure and only in that case can be
interpreted in terms of counterterms.

The basic assumption in the further discussion is that the
referred non renormalizable theory is finite and so the regularized
series can be summed up to some function which remains finite when
the regularization is removed.
This is just an assumption of mathematical consistency of the theory
--- if divergences are present even in exact solutions, then such
theory can not be regarded to be fundamental \cite{1}.
We seek for finite relations between physical quantities (these
quantities are
finite in terms of bare parameters if treated exactly and divergent in the
framework of perturbative
approach). We find that all physical quantities may be expressed as finite
coefficient series of some (finite number) physical expansion parameters.
(Remember that the ordinary renormalization procedure introduces infinite
number of new parameters).

We apply suggested method to Quantum
Gravity based on celebrated Einstein's classical Lagrangian. This
choice is motivated by our belief that this theory has much more
chance to be consistent then any other non renormalizable
theory known to us.

In Sec. 2 we briefly review conventional renormalization
procedure in a way that suits best for our purposes. Sec. 3 is
devoted to general description of the suggested method. In Sec.
4 application of the method to the abellian gauge field coupled
to gravity is described and in Sec. 5 we give some final
remarks and conclusions.

\section{Renormalization procedure}

Consider self-interacting scalar field $\phi$ with
some Lagrangian ${\cal L}(\phi,m_0,g_0)$ ($m_0$ and $g_0$ are the
bare mass
and coupling constant, respectively). The perturbation theory
produces diverging expressions for the Green's functions. So some
regularization is required. For definiteness let us work with
dimensional regularization \cite{6}. $n=4+2\epsilon$ is dimension of
the spacetime and $\mu$ is 't Hooft's dimensional parameter. After
regularization,
the {\cal S}-matrix elements $\sigma_i$ can be calculated (with the
help of LSZ reduction technique \cite{7}):
\begin{equation}
\sigma_i(g_0,m_0,p_k,\mu,\epsilon) =
\sum_{l}\sigma_i^l(m_0,p_k,\mu,\epsilon)g_0^l   \label{1}
\end{equation}
where $p_k$ are momenta and $\epsilon$
is the regularization parameter. In the limit $\epsilon\to 0$
coefficients $\sigma_i$ diverge. Let us introduce some functions:
\begin{equation}
m=\sum_{l}M_l(m_0,\mu,\Lambda,\epsilon)g_0^l      \label{2}
\end{equation}
\begin{equation}
g=\sum_{l}G_l(m_0,\mu,\Lambda,\epsilon)g_0^l    \label{3}
\end{equation}
Here $\Lambda$ is normalization point.
We can solve $g_0$ and $m_0$ from (\ref{2}) and (\ref{3}):
\begin{equation}
m_0=\sum_{l}M^*_l(m,\mu,\Lambda,\epsilon)g^l        ~\label{4}
\end{equation}
\begin{equation}
g_0=\sum_{l}G^*_l(m,\mu,\Lambda,\epsilon)g^l        ~\label{5}
\end{equation}
Now, substitute (\ref{4})-(\ref{5}) into (\ref{1}):
\begin{equation}
\sigma_i(g,m,p_k,\mu,\Lambda,\epsilon) =
\sum_{l}\sigma_i^{*l}(m,p_k,\mu,\Lambda,\epsilon)g^l   ~\label{6}
\end{equation}
If it is possible to choose the functions (\ref{2}) and (\ref{3})
in such a way that all divergences in (6) cancel, theory is
renormalizable. Of course if there exists one pair of functions
$m,g$ that satisfies this condition, the infinite number of such
pairs can be found and they are some finite functions of initial
$g$ and $m$, expandable in positive power series in
$g$ --- this is manifestation of the freedom in choosing
renormalization scheme.  The choice that seems natural, is
to take some physical quantities (e.g. the pole mass in (\ref{2}))
as $m$ and $g$. Of course quite often for technical reasons
other schemes are more convenient.

To reproduce the counter term technique of renormalization, let us
recall that LSZ technique prescribes to divide N-point Green's
functions by a factor $Z^{1/2}$ with $Z$ being the residue of the
propagator at the pole. We can define renormalized
field as $\phi_R=\phi Z_1^{-1/2}$ where $Z_1$ can differ from $Z$
by a finite multiplier. Now if we rewrite Lagrangian in terms of
$\phi_R$ and substitute instead of bare parameters their expansions
(\ref{4}) and (\ref{5}) we will recover Lagrangian with counter
terms. Although the described formulation of renormalization
procedure is fully equivalent to the counter term technique, for our
purposes it is more convenient.

Note that in this approach the substraction ambiguities are absent (in
fact they are fixed by the choice of renormalized parameters).

The feature of renormalization procedure that we want to underline
can be formulated as follows:
{\it After regularizing renormalizable theory it is enough just to
express all physical quantities in terms of few observables (their
number equals to the number of bare parameters) and divergences
will disappear.} In the next section we are going to demonstrate
that the same is true for nonrenormalizable theories too, with the
exception that the number of expansion parameters is more (but finite)
than that of bare ones.

\section{Renormalization of non renormalizable}

First of all let us formulate an assumption which will help us to
argument our method. We will assume that non renormalizable theory
under consideration is consistent --- i.e. nonperturbatively finite
and hence all divergences appearing in perturbative series are due to
the nonanalytic dependence of the expanded quantities on the bare
parameters.

The question we want to answer is whether it is possible
to extract any reliable information about the relations between
different physical quantities $\sigma_i$ even if they are given by
the series with divergent coefficients.
In order to see that sometimes the answer is `yes',
let us consider a simple mathematical example. Suppose we have
two functions $f_1$ and $f_2$ given by series with divergent
coefficients (we are interested in taking $\epsilon\to 0$ limit):
\begin{eqnarray}
f_1 &=& -\frac{g^3}{\epsilon}+
\frac{g^5}{\epsilon}+\frac{1}{2}\frac{g^5}{\epsilon^2}
+\cdots \nonumber\\
f_2 &=& 1+g+\frac{g^2}{\epsilon}-\frac{g^4}{\epsilon}
+\cdots                                             \label{7}
\end{eqnarray}
(If one expresses $g$ iteratively from the expression of $f_2$ and
substitutes into $f_1$, divergences do not cancel-``theory" is
not ``renormalizable")

Note that $k$-th inverse power of $\epsilon$ goes
together
with at least $k$-th power of $g^2$. Denoting $x\equiv g^2$ we can
rewrite (\ref{7}) as (in each term containing $\epsilon^{-k}$,
$g^{2k}$ is substituted by $x^k$):
\begin{eqnarray}
f_1 &=& -g\frac{x}{\epsilon}+
g^3\frac{x}{\epsilon}+
\frac{g}{2}\frac{x^2}{\epsilon^2}+
\cdots \nonumber\\
f_2 &=& 1+g+\frac{x}{\epsilon}-g^2\frac{x}{\epsilon}
+\cdots                                             \label{8}
\end{eqnarray}

Now let us for a moment consider $x$ as an independent parameter
and express iteratively $x$ from the second line in (\ref{8})
as power series in $g$ and $\alpha\equiv f_2-1-g$ (note that
the definition of $\alpha$ is automatically implied from
(\ref{8})) and substitute it into the expression of $f_1$.
It is easy to see that divergences disappear.
We get:
\begin{eqnarray}
x &=&\epsilon\left(\alpha+\alpha g^2+\cdots\right) \nonumber \\
f_1 &=& -(g\alpha-\frac{g}{2}\alpha^2+\cdots)        \label{9}
\end{eqnarray}
The right hand side of (\ref{9}) is the expansion of
\begin{equation}
f_1=-gln(1+\alpha)=-gln(f_2-g)       \label{9a}
\end{equation}
Indeed, we have obtained (\ref{7}) by `regularizing' and expanding
the following functions:
\begin{eqnarray}
f_1(g)=glng^2 & \to &
gln\frac{g^4/\epsilon+1}{g^2/\epsilon+1} \nonumber \\
f_2(g)=g+\frac{1}{g^2} & \to &
g+\frac{g^2/\epsilon+1}{g^4/\epsilon+1}       \label{10}
\end{eqnarray}
So we have recovered correct relation between $f_1$ and $f_2$ ---
(\ref{9a}) starting from series with divergent coefficients.
We considered this simple example to illustrate that the
diverging series  may as well contain some information
about relations between functions and this relations may be
extracted. It is worth mentioning that the method deals well with
different singular functions and reproduces correct series for
different 'regularizations'.

We would like to note, that although initially in (\ref{10}) we had
dependence over one parameter $g$, the expansion with finite
coefficients became possible only after introduction of one
extra expansion parameter $\alpha$. In fact parameters $g$ and
$\alpha$ are not independent.

The series in quantum field theory have a nice feature --- analogue
to the one that turned out useful in above example --- inverse powers
of $\epsilon$ always come together with at least some nonzero power
of coupling constant (bare or renormalizable). In other words any
renormalized Green's function or amplitude can be written as:
\begin{equation}
G_R=\sum_{i,k}f_{ik}g^i\left(\frac{g_R^\beta}{\epsilon}\right)^k
\label{11}
\end{equation}
Here coefficients $f_{ik}$ are finite in $\epsilon\to 0$ limit and
$\beta$ is determined by simple power counting.

Now it is clear how we can proceed in non renormalizable theory.
Consider some consistent non renormalizable theory with single
coupling constant. Acting along the lines of conventional
renormalization in the spirit described in previous section,
write expansion of e. g. pole mass $m$ in bare coupling and solve
from this expansion bare mass $m_0$. Also we can express bare
coupling $g_0$ from some physical amplitude or Green's function
(in the latter case renormalization of wave function is also
required) at some kinematics --- usually in renormalizable
theories it is an effective vertex $g_R$.

If the theory were renormalizable, then performing wave function
renormalization and inserting expressions of bare
parameters in $m$ and $g_R$  would make finite any
Green's function.
In non renormalizable theory we are left with
series for Green's functions that still contain divergences.

Next introduce in (\ref{11}) $x$ instead of $g_R^\beta$ and
express it from any convenient Green's function or amplitude
as series in $g_R$ and $\alpha$ (where definition of $\alpha$
would be automatically implied just like in the example above).
Evidently, inserting this
series into any other Green's function will lead to series free
of divergences. The price we have to pay for it
is introduction of an extra expansion parameter. Of course $g_R$ and
$\alpha$ are not independent. We do not know whether the relation
between them can be established perturbatively.

Due to our main assumption, nonperturbative solutions of the theory are
finite in terms of bare parameters. In our approach we do not do any
substractions and do not introduce counter terms, so the associated
ambiguities are absent.

 Of course the status of final series will
depend on the theory under consideration --- hopefully, for consistent
theories they will not be worse than asymptotic.

The method can be generalized for the case of several
bare couplings avoiding introduction of more then one
extra expansion parameter.

So the suggested method coincides with the ordinary renormalization
procedure for renormalizable theories and implies introduction of an
extra effective parameter for non renormalizable ones.
Formally it works not only for non renormalizable theories which are finite
outside perturbation theory but, unfortunately, it will produce
series with finite coefficients for the theories where infinities are
present even in exact solutions.

The described
method is easily applied within the framework of any
regularization where divergences appear only as powers of some
regulator. For other regularizations more (but finite number) of
extra expansion parameters will be required.

\section{Application to Quantum Gravity}

Let us illustrate the general ideas presented in previous section on
the example of Quantum Gravity. The non renormalizability of
Einstein's gravity coupled to scalar field and to fermion or photon
fields was demonstrated in \cite{ghooft} and \cite{deser},
respectively. We will consider the latter case --- photon field.
First let us derive the Feynman
rules (we follow ref. \cite{Capper}).
The Lagrangian density has the form:
$$
{\cal L}={\cal L}_G+{\cal L}_A
$$
where ${\cal L}_G$ is the familiar Einstein Lagrngian:
$$
{\cal L}_G=\frac{2}{k^2}\sqrt{-g}g^{\mu\nu}R_{\mu\sigma\nu}^\sigma
$$
with $g^{\mu\nu}$ being the metric tensor and $R$ the curvature
tensor. ${\cal L}_A$ denotes the generally covariant photon Lagrangian,
defined by minimal substitution
$$
{\cal L}_A=-\frac{1}{4}\sqrt{-g}g^{\mu\nu}g^{\alpha\beta}
F_{\mu\nu}F_{\alpha\beta}
$$
with $F_{\mu\nu}=\partial_\mu A_\nu-\partial_\nu A_\mu $. (We
work in Euclidian space.)
Defining ${\cal G}^{\mu\nu}\equiv \sqrt{-g}g^{\mu\nu}$ we can rewrite
${\cal L}_G$ and ${\cal L}_A$ in arbitrary $n$ dimensions as
$$
{\cal L}_G=\frac{1}{2k^2}\left(
{\cal G}^{\rho\sigma}{\cal G}_{\lambda\mu}{\cal G}_{\kappa\nu}-
\frac{1}{n-2}
{\cal G}^{\rho\sigma}{\cal G}_{\mu\kappa}{\cal G}_{\lambda\nu}-
2\delta^\sigma_\kappa\delta^\rho_\lambda {\cal G}_{\mu\nu}\right)
\partial_\rho {\cal G}^{\mu\kappa}{\cal G}_\sigma^{\lambda\nu}
$$
$$
{\cal L}_A=-\frac{1}{4}
\left(-det {\cal G}\right)^{-\frac{1}{n-2}}
{\cal G}^{\mu\nu}{\cal G}^{\alpha\beta}F_{\mu\alpha}F_{\nu\beta}
$$
where $\delta$-s denote $n$-dimensional Kronecker symbols.
We can write the generating functional as follows
\begin{eqnarray}
Z[{\cal J}_{\mu\nu},J_\mu]  & = & \int {\cal D}({\cal G}^{\mu\nu})
{\cal D}(A^{\mu})\Delta[{\cal G}^{\mu\nu},A_\alpha]  \nonumber \\
 & & \delta(\partial_\mu {\cal G}^{\mu\nu})
 \delta(\partial_\alpha A^\alpha)
e^{i\int d^nx ({\cal L}_G+{\cal L}_A)+J_\mu A^\mu+
{\cal J}_{\mu\nu}{\cal G}^{\mu\nu})} \nonumber
\end{eqnarray}
$$
\Delta[{\cal G}^{\mu\nu},A_\alpha] \int {\cal D}\Omega
\delta(\partial_\mu {\cal G}_\Omega^{\mu\nu})
\delta(\partial_\alpha A^{\alpha \Omega})=1
$$
Defining graviton field $\phi^{\mu\nu}$ by
$k\phi^{\mu\nu}={\cal G}^{\mu\nu}-\delta^{\mu\nu}$ we can expand
${\cal G}_{\mu\nu}$ as
$$
{\cal G}_{\mu\nu}=\delta_{\mu\nu}-k\phi_{\mu\nu}+
k^2\phi_{\mu\lambda}\phi_{\lambda\nu}+
k^3\phi_{\mu\alpha}\phi_{\alpha\beta}\phi_{\beta\nu}+O(k^4)
$$
(We work in Euclidian space so there is no need to distinguish
between upper and lower indices of $\phi$). In terms of graviton
field Lagrangian takes the form:
$$
{\cal L} = \sum_{i=2}^{\infty}k^{i-2}{\cal L}_{(i)} =
\sum_{i=2}^{\infty}k^{i-2}
\left({\cal L}_{G(i)}+{\cal L}_{A(i)}\right)
$$
To define graviton and photon propagators we need quadratic parts:
$$
{\cal L}_{G(2)}=\frac{1}{2}\partial_\mu \phi_{\nu\lambda}
\partial_\mu \phi_{\nu\lambda}-\frac{1}{2(2-n)}
\partial_\mu \phi_{\nu\nu}\partial_\mu \phi_{\lambda\lambda}-
\partial_\mu \phi_{\mu\nu}\partial_\rho \phi_{\rho\nu}
$$
$$
{\cal L}_{A(2)}=-\frac{1}{4}F^{\mu\nu}F^{\mu\nu}
$$
The ghost propagator is defined from the expression (note that using
particular type of gauge $\partial_\alpha A_\alpha=0$ no ghost
corresponding to the photon field is required):
\begin{eqnarray}
\Delta[{\cal G}^{\mu\nu},A_\alpha]
 & = & \int {\cal D}(\zeta_\lambda)
{\cal D}(\eta_\nu) exp
\Biggl\{i\int d^nx \eta_\nu\biggl[\delta_{\mu\nu}\partial^2-k\Bigl(
\partial_\lambda\partial_\mu\phi_{\mu\nu} - \nonumber \\
 & - &
\phi_{\mu\rho}\delta_{\nu\lambda}
\partial_\rho\partial_\mu
-\partial_\mu\phi_{\mu\rho}
\delta_{\nu\lambda}\partial_\rho+\partial_\mu\phi_{\mu\nu}
\partial_\lambda\Bigr)\biggr]\zeta_\lambda\Biggr\}  \label{12}
\end{eqnarray}
Accordingly propagators have the following form: \\
photon propagator
$$
D_{\mu\nu}(q^2)=\frac{1}{q^2}
\left(-\delta_{\mu\nu}+\frac{q_\mu q_\nu}{q^2}\right)
$$
ghost propagator
$$
\Delta_{\mu\nu}=\frac{\delta_{\mu\nu}}{q^2}
$$
graviton propagator
\begin{eqnarray}
{\cal D}_{\alpha\beta,\lambda\nu}(p) &=&
\frac{1}{2p^2}(\delta_{\lambda\alpha}\delta_{\beta\mu}+
\delta_{\alpha\mu}\delta_{\beta\lambda}-
2\delta_{\alpha\beta}\delta_{\lambda\mu}) - \nonumber \\
 &-&\frac{1}{2p^4}(p_\lambda p_\alpha\delta_{\mu\beta}+
p_\mu p_\alpha\delta_{\lambda\beta}+p_\lambda p_\beta
\delta_{\mu\alpha}+p_\mu p_\beta\delta_{\lambda\alpha})
+ \nonumber \\
 &+&\frac{1}{p^4}(p_\lambda p_\mu\delta_{\alpha\beta}+
p_\alpha p_\beta\delta_{\lambda\mu}) \nonumber
\end{eqnarray}
Vertices are defined from the ${\cal L}_{(i>2)}$ and also from
$\eta\phi\zeta$ terms in (\ref{12}).
It is easy to see that any $N$-particle vertex has the factor
$k^{N-2}$ --- this fact is important for our further analysis.

Consider a Feynman diagram containing $N_i$ $i$-particle (graviton
or photon) vertices,
with $E$ external legs and $I$ internal ones. It is straightforward
to relate these quantities to the number of loop integrations $l$:
$$
l=\frac{1}{2}(\sum_{i}N_i(i-2)-E+2)
$$
If we use dimensional regularization, then $l$ loop integrations may
produce at most $(\frac{1}{\epsilon})^l$ divergence. As we have
mentioned above any $i$-particle vertex has the order $k^{(i-2)}$,
so comparing powers of $k$ and $\frac{1}{\epsilon}$ we see that any
$N$-point Green's function can be written as
$$
G_N=k^{N-2}\sum_{m,n}
\left(\frac{k^2}{\epsilon}\right)^mk^nC_{m,n}
$$
where coefficients $C_{m,n}$ are free of divergences.

The dimensional regularization is unique up to arbitrary normalization
function $f(n)$, which should satisfy only one condition $f(4)=(2\pi)^{-4}$
\cite{3}.
This arbitrary function $f(n)$ produces ambiguities which can be included
into definition of renormalized parameters in renormalizable theories and
they make problems in non renormalizable theories. However, if the exact
solutions of the theory are finite, then ambiguities are formal
expansions of zero and they should be dropped. Indeed N-point Green's
function's expressions have the form:
$$
G_N=k^{N-2}\sum (k^2f(n))^lC_l
$$
in this expression $l$ is number of loops. As far as $G_N$ is finite in
terms of $k$, it is evident that in the $n\to 4$ limit $G_N$ depends only
on $f(4)$. Note that on the grounds of the same arguments one should  drop
the $\mu$-dependent terms.

Let  us consider amputated Green's function at symmetric point $q^2$: $$
\Gamma_{\alpha\beta\sigma\lambda}\equiv
<{\cal G}_{\alpha\beta}A_\sigma A_\lambda>
$$
After performing wave function renormalization it takes the form:
\begin{eqnarray}
\Gamma_{\alpha\beta\sigma\lambda} &\sim &
\delta_{\alpha\beta}\delta_{\lambda\sigma}C_1(q^2)+
(\delta_{\alpha\lambda}\delta_{\beta\sigma}+
\delta_{\alpha\sigma}\delta_{\beta\lambda})C_2(q^2)+
\delta_{\alpha\beta}p_\sigma'p_\lambda C_3(q^2) + \nonumber \\
&+&\delta_{\lambda\sigma}(p_\alpha p_\beta'+p_\alpha'p_\beta)
C_4(q^2) +   \nonumber \\
 & + & (\delta_{\alpha\sigma}p_\lambda p\beta'+\delta_{\beta\sigma}
p_\lambda p_\alpha'+\delta_{\alpha\lambda}p_\beta p_\sigma'+
\delta_{\beta\lambda}p_\alpha p_\sigma')C_5(q^2) \label{13}
\end{eqnarray}
We suppose that there exist some finite exact formfactors $C_i$ that
stand in (\ref{13}) as coefficients of independent tensor
structures.
The perturbative expansion gives:
\begin{eqnarray}
2(2-n)C_1 & = & kq^2+k^3q^4\left(\frac{a_1}{\epsilon}+a_2+
a_3(\epsilon)\right) + \nonumber \\
 & + & k^5q^6\left(\frac{a_4}{\epsilon^2}+
\frac{a_5}{\epsilon}+a_6+a_7(\epsilon)\right)+\cdots
\label{14}
\end{eqnarray}
\begin{equation}
C_2=\frac{kq^2}{2}+k^3q^4
\left(\frac{b_1}{\epsilon}+b_2+b_3(\epsilon)\right)+
k^5q^6
\left(\frac{b_4}{\epsilon^2}+\frac{b_5}{\epsilon}+b_6+
b_7(\epsilon)\right)+\cdots \label{15}
\end{equation}
Let us introduce renormalized coupling $k_R$ as:
\begin{equation}
k_R=\frac{2(2-n)C_1(\Lambda^2)}{\Lambda^2} \label{16}
\end{equation}
here $\Lambda$ is normalization point.
Solving (\ref{14}) iteratively for $k$ we obtain:
\begin{eqnarray}
k & = & k_R-k_R^3\Lambda
\left(\frac{a_1}{\epsilon}+a_2+a_3(\epsilon)\right) + \nonumber \\
 & + & k_R^5\Lambda^4\left(\frac{3a_1^2-a_4}{\epsilon^2}+
\frac{6a_1a_2-a_5}{\epsilon}+\tilde a(\epsilon)\right)+\cdots
\label{*}
\end{eqnarray}
Inserting this expansion for bare coupling into expression of
$C_2$ --- (\ref{15}) we find:
\begin{eqnarray}
C_2 &=& \frac{k_Rq^2}{2}+k_R^3\left(\frac{k_R^2}{\epsilon}
\left(b_1q^4-\frac{a_2}{2}q^2\Lambda^2\right)+
k_R^2\tilde b_2\right) + \nonumber \\
&+& k_R\left(\frac{k_R^2}{\epsilon}\right)^2
\left(b_4q^6-3a_1b_1q^4\Lambda^2-
\frac{3a_1^2-a_4}{2}q^2\Lambda^4\right)  + \nonumber \\
&+&k_R^3\frac{k_R^2}{\epsilon}\left(b_5q^6-
3(b_1a_2+b_2a_1)q^4\Lambda^2+
\frac{6a_1a_2-a_5}{2}q^2\Lambda^4\right)+\cdots \nonumber
\end{eqnarray}
Consider extension of function $C_2$ to two parameter dependence by
replacing ${k_R\over\epsilon}\to {x\over\epsilon}$:
\begin{eqnarray}
C_2^*(q^2) &=& k_R\frac{q^2}{2}
+k_R\left(\frac{x}{\epsilon}
\left(b_1q^2-\frac{a_1}{2}q^2\Lambda^2\right)+k_R^2\tilde b_2
\right) + \nonumber \\
&+& k_R\left(\frac{x}{\epsilon}\right)^2\left(b_4q^2-3a_1b_1q^2\Lambda^2+
\frac{3a_1^2-a_4}{2}q^2\Lambda^4\right) + \nonumber \\
&+& k_R^3\frac{x}{\epsilon}\left(b_5q^6-3\left(b_1a_2+b_2a_1\right)
q^4\Lambda^2+\frac{6a_1a_2-a_5}{2}q^2\Lambda^4\right) + \nonumber \\
 & + & k_R^5\tilde b_6+\cdots \label{**}
\end{eqnarray}
Of course, taking $x=k_R^2$ we recover $C_2$ from $C_2^*$.
Let us denote the sum of diverging terms in (\ref{**}) by
$\tilde \alpha$
(according to our assumption $\tilde \alpha$ is nonperturbatively finite).
\begin{eqnarray}
\tilde \alpha(q^2)
&=& k_R\frac{x}{\epsilon}\left(b_1q^4-\frac{a_1}{2}
q^2\Lambda^2\right) + \nonumber \\
 & + &k_R\left(\frac{x}{\epsilon}\right)^2
\left(b_4q^6-3q^4\Lambda^2a_1b_1+
\frac{3a_1^2-a_4}{2}q^2\Lambda^4\right) + \nonumber \\
&+& k_R^3\frac{x}{\epsilon}\left(b_5q^6-
3\left(b_1a_2+b_2a_1\right)q^4\Lambda^2+
\frac{6a_1a_2-a_5}{2}q^2\Lambda^4\right)+\cdots = \nonumber \\
&=& C_2^*-k_R\frac{q^2}{2}-k_R^3\tilde b_2-k_R^5\tilde b_6+\cdots
\label{17}
\end{eqnarray}
Now solve (\ref{17}) iteratively for $x$ as series in
\begin{equation}
\alpha(\Lambda^2)=
\frac{\tilde \alpha(\Lambda^2)}{k_R\left(b_1\Lambda^4-
\frac{a_1}{2}\Lambda^4\right)} \label{18}
\end{equation}
\begin{eqnarray}
x &=& \epsilon\alpha(\Lambda^2)-\epsilon k_R\alpha^2(\Lambda^2)
\frac{b_4-3a_1b_1+\frac{3a_1^2-a_4}{2}}{b_1-\frac{a_1}{2}}\Lambda^2
- \nonumber \\
&-&\epsilon k_R^3\alpha(\Lambda^2)
\frac{b_5-3(b_1a_2+b_2a_1+\frac{6a_1a_2-a_5}{2})}{b_1-\frac{a_1}{2}}
\Lambda^2+\cdots \label{19}
\end{eqnarray}
(Of course we could solve (\ref{17}) at some other
$\Lambda_1\neq\Lambda$).
The defined effective coupling constants (\ref{16}) and (\ref{18})
will enable us to make finite coefficients of any Green's function (or
physical amplitude).
Indeed, after renormalization of wave function and coupling constant any
Green's function takes form:
\begin{equation}
G_N=k_R^{N-2}\sum \left( {k_R^2\over\epsilon}\right)^mk_R^nG_{mn}
\label{n}
\end{equation}
We introduce `related' extension of (\ref{n}):
$$
G_N^*=k_R^{N-2}\sum \left( {x\over\epsilon}\right)^mk_R^nG_{mn}
$$
and substitute (\ref{19}) for $x$. In such manner we get series of
$\alpha$ and $k_R$ with finite coefficients.
In this manner we can express {\it any} Green's function as finite
coefficient series of two effective parameters (these can be defined
from some quantities other then formfactors $C_{1,2}$). We do not
present explicit calculations of particular processes here.

\section{Conclusions}

So we have described a method of `renormalization' of perturbative
series in nonrenormalizable theories. For renormalizable ones it
just coincides with the usual renormalization procedure. The method
is based on the assumption of non perturbative finiteness of exact
solutions of the theory in terms of bare parameters and on the suitable
introduction of an extra effective expansion
parameter (which is not independent). We are unable to find relation
between effective couplings within the framework of perturbation
theory, so for numeric analysis we need one more experimental value
then the number of bare couplings and masses.

The method can be applied to any theory. (E.g. for standard model
plus gravity one may use suitably adjusted extra effective
expansion parameter defined in previous section.)

Unfortunately in full analogy to the conventional
renormalization procedure for renormalizable theories, suggested
method is insensitive to the consistency of the theory --- it will
formally produce order by order finite series even for inconsistent
theories.
So establishing of asymtotic character of final series in given
theory is desirable,
but unfortunately this problem is too complicated
(e.g. it is not completely solved even for so `well explored' theory
as QED). Although we hope that quantum gravity based on Einstein's
Lagrangian is nonperturbatively finite and calculations using
suggested method are meaningful.

We do not introduce counter terms and do not do substractions and so do
not have associated ambiguities. Ambiguities, which are introduced by the
regularization itself are formal expansions of zero and should be dropped.

Although we used the dimensional regularization it is possible to apply
suggested method to other regularizations.

Acknowledgements: We would like to thank G. Jorjadze for useful
discussions. One of us (J.G.) wants to thank also D. Broadhurst
for interest to this work.

\end{document}